\documentclass[a4,center,fleqn]{NAR}
\usepackage{amsmath}
\usepackage{amsthm}
\usepackage{graphicx}
\usepackage{color}

\usepackage{NAR-natbib}

\newtheorem{defn}{Definition}

\DeclareMathOperator*{\argmin}{arg\,min}

\copyrightyear{}
\pubdate{}
\pubyear{}
\jvolume{}
\jissue{}

\begin{document}

\title{Genetic robustness of let-7 miRNA sequence-structure pairs}

\author[Qijun He \textit{et~al}.]
       {Qijun He\,$^{\text{\sfb 1}}$, Fenix W. Huang\,$^{\text{\sfb 1}}$, Christopher Barrett\,$^{\text{\sfb 1,4}}$
         and Christian M. Reidys\,$^{\text{\sfb 1,2,3}*}$
         \footnote{To whom correspondence should be addressed.
Tel: +01 540 231-2317; Email: duckcr@bi.vt.edu}}
         
\address{
$^{\text{\sf 1}}$ Biocomplexity Institute of Virginia Tech, Blacksburg, VA, USA.
$^{\text{\sf 2}}$ Department of Mathematics, Virginia Tech, Blacksburg, VA, USA.
$^{\text{\sf 3}}$ Thermo Fisher Scientific Fellow in Advanced Systems for Information Biology.
$^{\text{\sf 4}}$ Department of Computer Science, Virginia Tech, Blacksburg, VA, USA.
}

%and
%$^{2}$Affiliation of Both Co-Authors}
% Affiliation must include:
% Department name, institution name, full road and district address,
% state, Zip or postal code, country

\history{%
}

\maketitle

\begin{abstract}
%200 words
  Genetic robustness, the preservation of evolved phenotypes against genotypic mutations, is one of the central concepts in
  evolution. In recent years a large body of work has focused on the origins, mechanisms, and consequences of robustness in
  a wide range of biological systems. In particular, research on ncRNAs studied the ability of sequences to maintain folded
  structures against single-point mutations. In these studies, the structure is merely a reference. However, recent work revealed
  evidence that structure itself contributes to the genetic robustness of ncRNAs. We follow this line of thought and
  consider sequence-structure pairs as the unit of evolution and introduce the spectrum of inverse folding rates (IFR-spectrum)
  as a measurement of genetic robustness. Our analysis of the miRNA let-7 family captures key features of
  structure-modulated evolution and facilitates the study of robustness against multiple-point mutations. 
\end{abstract}

\section{Introduction}
%1.5 page introduction
Genetic robustness can be characterized in terms of the variation of phenotype distribution induced by genotypic change
\cite{gu2003role, de2003perspective, schlichting1998phenotypic} and concerns the insensitivity of a phenotype to
genetic changes. Mutational robustness has been studied in the context of noncoding RNA (ncRNA)
\cite{borenstein2006direct, rodrigo2012describing}.
RNA consists of a single strand of nucleotides (A,C,G,U) that can fold and bond to itself through base pairing. ncRNAs
are known to function in aptamer binding as riboswitches, in chemical catalysis as ribozymes and in RNA splicing such
as spliceosome \cite{darnell2011rna, breaker1996engineered, serganov2007ribozymes, breaker1994inventing}. Most importantly,
it is the folded structure that is underlying all these mechanisms, allowing for the interaction with and subsequent
modification of other biological molecules. The self folding of RNA makes it an ideal object to study genotype-phenotype relations.
The well-established energy based prediction of RNA secondary structure, a $2$-dimentional coarse grain of the real three
dimensional structure, makes such studies feasible \cite{Waterman:78a, Mathews:99, Zuker:81, Hofacker:94a}.

Structures are an important determinant of the function of ncRNAs, whence the robustness of an RNA can be characterized
in terms of the variation of secondary structure distribution, i.e.~the stability of a secondary structure in the face of 
genetic sequence changes. Structural robustness of ncRNAs is considered to be a key component of the fitness of the molecule and
much research has been conducted to identify the footprints of natural selection on secondary structures of
sncRNAs \cite{borenstein2006direct, rodrigo2012describing}. In \cite{borenstein2006direct}, Borenstein and
Ruppin define neutrality of an RNA sequence $\sigma=a_1a_2...a_n$ by $\eta(\sigma)=1-\langle d\rangle/n$, where $\langle d\rangle$ denotes the
average, taken over all $3n$ single-point mutants of $\sigma$, of the base pair distance $d$ between the minimum free energy
(MFE) structure, $S_0$, of $\sigma$ and the MFE structures of single-point mutants. The RNA sequence, $\sigma$, is
then defined to be robust if $\eta(\sigma)$ is greater than the average neutrality of 1000 control sequences generated by
the program RNAinverse \cite{lorenz2011viennarna}, which fold into the same target structure, $S_0$. The main finding of
\cite{borenstein2006direct} is that precursor miRNAs (pre-miRNA) exhibit a significantly higher level of mutational robustness
than random RNA sequences, having the same structure. Subsequently Rodrigo et al. \cite{rodrigo2012describing} undertook a
similar analysis for bacterial small RNAs. Their main finding was, that, surprisingly, bacterial sncRNAs are not significantly
more robust when compared with 1000 sequences having the same structure, as computed by RNAinverse.
\cite{rodrigo2012describing} based their findings on the notion of \emph{ensemble diversity} defined earlier in
\cite{gruber2008strategies}. 

In the above mentioned papers, robustness is defined by taking the average of all $3n$ single-point mutants, the underlining
assumption being, that all $3n$ single-point mutations are equally likely to occur and, in addition, taking exclusively single-point mutations
into account. Incorporation of multiple-point mutations poses obvious difficulties, since the number of sequences that need to be
taken into consideration will grow exponentially.

\enlargethispage{-65.1pt}

The neutral theory of Motoo Kimura \cite{kimura1983neutral}, stipulates that evolution is achieved by neutral mutations, that is, by
mutations, that are necessarily compatible with the underlying structure. This is in accordance with recent findings showing
that secondary structures have a \emph{genuine} influence on selection in RNA genes. In \cite{mimouni2008analysis}, Hein {\it et al.} classify
stem positions into structural classes and validate that they are under different selective constraints.
In \cite{price2011neutral}, the authors observe neutral evolution in Drosophila miRNA and evolution increasing the thermal
dynamic stability of the RNA. \cite{pollard2006rna, beniaminov2008distinctive} study Human Accelerated Regions (HAR) in brains of primates, i.e.~noncoding RNAs
with an accelerated rate of nucleotide substitutions along the lineage between human and chimpanzee. \cite{beniaminov2008distinctive} concludes that
this increased rate of nucleotide substitutions is of central importance for the evolution of the human brain.
The most divergent of these regions, HAR1, has been biochemically confirmed to fold into distinct RNA secondary structures
in human and chimpanzee. Interestingly,
the mutations in the human HAR1 sequence, compared to the chimpanzee sequence, stabilize their respective RNA structure \cite{pollard2006forces},
suggesting a shape modulated evolution.

% **************************************************************
% Keep this command to avoid text of first page running into the
% first page footnotes
%\enlargethispage{-65.1pt}
% **************************************************************

Recently, \cite{barrett2017sequence} proposed a framework considering 
%\enlargethispage{-65.1pt}
RNA sequences and their RNA secondary structures
simultaneously. This gives rise to an information theoretic framework for RNA sequence-structure pairs. In particular, the
authors studied the ``dual Boltzmann distribution'', i.e.~the Boltzmann distribution of sequences with respect to a fixed
structure. The authors develop a Boltzmann sampler of sequences with respect to a given structure and study the ``inverse
folding rate'' (IFR) of the sampled sequences, i.e.~the proportion of the sampled sequences whose MFE structure equals to the
given structure. \cite{barrett2017sequence} reports that natural structures have higher IFR than random structures, suggesting that natural
structures have higher intrinsic robustness than random structures.

This suggests an approach to consider sequences and structures-as {\it pairs}-as the unit of evolution, instead of just
sequences in isolation. In this paper, we generalize the notion of {\it inverse folding rate} (IFR) and introduce a novel
profile, the IFR-{\it spectrum}, of a sequence-structure pair. By construction, this spectrum entails both sequence and
structure information.
The key idea is that mutations will be biased by the underlying thermodynamic energy of the respective structure, instead of
being equally likely to occur.

We shall conduct a detailed study on the let-7 family miRNAs, small endogenous noncoding RNAs,
that regulate the expression of protein coding genes in animals.
The short, mature miRNAs ($\sim$22 nt) originate from longer RNA precursor molecules that fold into a stem-loop hairpin structure.
The secondary structure of miRNA stem-loops serves a crucial role in the miRNA gene maturation process \cite{lee1993c}. The stem-loop
structure has been under evolutionary pressures to conserve its structure. Such stabilizing pressures favor robust configurations
and may have led to the evolution of robust structures. Furthermore, the let-7 miRNA family has been widely detected in metazoans,
ranging from human to fruit fly \cite{tanzer2004molecular}. These features make the let-7 gene family a particularly suitable test bed for studying the
evolution of genetic robustness.

We organize our study as follows: first we shall extend the analysis of \cite{borenstein2006direct} to IFR-spectra and observe
that most native sequence-structure pairs have a higher IFR-spectrum than sequence-structure pairs obtained by the inverse
folding algorithm. We shall investigate different aspects of the robustness of IFR-spectra of native sequence-structure
pairs and show that these are distinctively more robust against multiple-point mutations. Secondly, we conduct cross-species
comparisons of native sequence-structure pairs, observing that higher metazoan species have higher IFR-spectra. Our analysis
suggests that IFR-spectra are being increased in the course of shape-modulated evolution.

%%%
%%%%%%%%%%%%%%%%%%%%%%%%%%%%%%%%%%%%%%%%%%%%%%%%%%%%%%%%%%%%%%%%%%%%%%%%%%%%%%%%%%%%%%%%%%%%%%%%%%%%%%%%%%%%%%%%%%%%%%%%%%%%%%
%%%
\section{MATERIALS AND METHODS}
%%%
%%%%%%%%%%%%%%%%%%%%%%%%%%%%%%%%%%%%%%%%%%%%%%%%%%%%%%%%%%%%%%%%%%%%%%%%%%%%%%%%%%%%%%%%%%%%%%%%%%%%%%%%%%%%%%%%%%%%%%%%%%%%%%
%%%
%2.5 pages 

\subsection{IFR-spectrum: a sequence-structure pair profile.}

In this section we introduce the technical details of our framework, starting with Borenstein and Ruppin's definition
of neutrality \cite{borenstein2006direct}, as the average of all $3n$ single-point mutants. By construction, all mutations
are equivalent and structure has no
  influence on the mutation rate.
  However, as suggested in \cite{mimouni2008analysis,price2011neutral,pollard2006forces}, structure genuinely affects mutation
  rate and mutations in turn further increase the thermal dynamic stability of the structure.
  We consider here mutations with respect to a \emph{fixed} structure and consequently deal
  exclusively with \emph{compatible} mutations. Furthermore, the idea of the following is to favor energetically beneficial
  mutations, while penalizing detrimental mutations.
To adequately quantify the above, we revisit some of the basic concepts of the thermodynamic model of RNA secondary
structure.

The free energy $\eta$ can be considered as a result of pairing sequences and structures as follows:
$$
\eta: \mathcal{Q}_4^n \times \mathcal{S}_n \to \mathbb{R} \sqcup +\infty,
$$
where $\mathcal{Q}_4^n$ and $\mathcal{S}_n$ denote the space of sequences, $\sigma$, and the space of secondary structures,
$S$, respectively and $\eta(\sigma,S)$ is the energy of $S$ on $\sigma$. This mapping is computed as the sum of the
  energy contributions of individual base-pairs \cite{Nussinov:78}. A more elaborate model \cite{Mathews:99, Turner:10}
  evaluates the total free energy to be the sum of from the energies of loops involving multiple base-pairs.
We remark that, if a sequence, $\sigma$, is not compatible with a structure, $S$, then $\eta(\sigma,S)=+\infty$.
Then MFE-folding is a map from sequences to distinguished structures, i.e.~the minimum free energy structures:
$$
\text{mfe}\colon \mathcal{Q}_4^n \ \to \  \mathcal{S}_n, \quad \sigma \ \mapsto \ \argmin_{S\in \mathcal{S}_n}  \eta(\sigma,S).
$$
To incorporate the thermodynamic information into the probability of mutations, we introduce the notion of the Boltzmann
distribution of $k$-point mutants of a sequence-structure pair. To this end, we consider the partition function
of $k$-point mutants of a sequence-structure pair, namely, the partition function of all sequences that are at Hamming
distance $k$ to the given sequence with respect to the given structure. 

%%%
%%%%%%%%%%%%%%%%%%%%%%%%%%%%%%%%%%%%%%%%%%%%%%%%%%%%%%%%%%%%%%%%%%%%%%%%%%%%%%%%%%%%%%%%%%%%%%%%%%%%%%%%%%
%%%
\begin{defn}
  Let $\sigma$ be a sequence of $n$ nucleotides and let $S$ be its associated structure, i.e.~its MFE- or native
  structure. 
  Then the partition function of $k$-point mutants of $\sigma$ with respect to $S$ is given by:
\begin{equation}\label{E:KL}
 \qquad \qquad \qquad \quad Q_{\sigma,k}^{S}= \sum_{\sigma',h(\sigma,\sigma')=k} e^{-\frac{\eta(\sigma',S)}{RT}},
\end{equation}
where $h$ is the Hamming distance, $\eta(\sigma',S)$ is the energy of $S$ on $\sigma'$, $R$ is the universal gas constant
and $T$ is the temperature.
\end{defn}
%%%
%%%%%%%%%%%%%%%%%%%%%%%%%%%%%%%%%%%%%%%%%%%%%%%%%%%%%%%%%%%%%%%%%%%%%%%%%%%%%%%%%%%%%%%%%%%%%%%%%%%%%%%%%%
%%%
Eq.~(\ref{E:KL}) represents the ``dual'' of McCaskill's partition function \cite{mccaskill1990equilibrium} with an additional Hamming
distance filtration \cite{huang2017efficient}. Given this partition function, we are in position to introduce the
\emph{Boltzmann distribution} of $k$-point mutants of $\sigma$ with respect to $S$: the probability of a specific
$k$-point mutant $\sigma^*$ of $\sigma$, $h(\sigma^*,\sigma)=k$, with respect to $S$:
$$
Pr_{\sigma,k}^S(\sigma^*)=\frac{e^{-\frac{\eta(\sigma^*,S)}{RT}}}{Q_{\sigma,k}^{S}}.
$$
This expression allows us to consider Boltzmann weighted mutations, taking into account the free energy, when
realizing $S$.

In order to quantify mutational robustness, i.e.~the ability to maintain the structure, we consider all $k$-point
mutants, that fold again into $S$. We call the fraction of $k$-point mutants, $\sigma^*$, that fold into $S$, the
\emph{inverse folding rate} (IFR) of the sequence structure pair $(\sigma,S)$ at $k$, that is, we have
$$
\text{IFR}_{\sigma,k}^S= \sum_{\text{mfe}(\sigma^*)=S} Pr_{\sigma,k}^S(\sigma^*).
$$
$\text{IFR}_{\sigma,k}^S$ thus quantifies the mutational robustness of a sequence-structure pair, $(\sigma,S)$, with respect to
Boltzmann weighted $k$-point mutations, taking into account the energy of the sequence, when assuming the structure $S$.

$\text{IFR}_{\sigma,1}^S$ can be viewed as a variation of Borenstein and Ruppin's definition of neutrality.
Instead of a uniform distribution of all $3n$ single-point mutants, a Boltzmann weighted distribution for these mutants
is employed. In contrast to using base pair distance as the metric on structure space, we restrict ourselves to a discrete
metric.

The \emph{$\text{IFR}_{\sigma,k}^S$-spectrum} of a sequence-structure pair, $(\sigma,S)$, i.e.~the collection of $\text{IFR}_{\sigma,k}^S$ for
varying $k$, allows us to study multiple-point mutations instead of confining the analysis to single-point mutations. In this
study, we shall analyze the $\text{IFR}_{\sigma,k}^S$-spectrum of sequence-structure pairs of Hamming distances $1\le k\le 20$.

The $\text{IFR}_{\sigma,k}^S$-spectrum can be viewed as the conditional probability of the IFR in \cite{barrett2017sequence},
conditional to specific Hamming distances. As a result, we have:
$$
\text{IFR}^S=\sum_{k=0}^n Pr_{\sigma}^S(k) \text{IFR}_{\sigma,k}^S,
$$
where $Pr_{\sigma}^S(k)=\frac{Q_{\sigma,k}^{S}}{Q^S}$, $Q^S= \sum_{\sigma'} e^{-\frac{\eta(\sigma',S)}{RT}}$.

\subsection{The $\text{IFR}_{\sigma,k}^S$-spectrum via Boltzmann sampling.}
Computing $\text{IFR}_{\sigma,k}^S$ strictly requires folding all $\sigma'$ that are $S$-compatible, such that
$h(\sigma,\sigma')=k$. While this task is impractical for large $k$, $\text{IFR}_{\sigma,k}^S$ can be efficiently computed
by means of Boltzmann sampling. This is conducted here via the Hamming Distance Restricted Dual Sampler (HRDS)
\cite{huang2017efficient}, which facilitates the approximation of $\text{IFR}_{\sigma,k}^S$ for any given sequence-structure
pair, $(\sigma,S)$, and Hamming distance, $k$. HDRS takes as input $(\sigma,S)$ and $k$ and outputs sequences, $\sigma^*$,
having Hamming distance $k$ with probability $\frac{e^{-\frac{\eta(\sigma^*,S)}{RT}}}{Q_{\sigma,k}^{S}}$. We then have: 
$$
\text{IFR}_{\sigma,k}^S\approx \frac{\# \text{ of sequences folding into S}}{\# \text{ of sampled sequences}}.
$$
In this paper, $\text{IFR}_{\sigma,k}^S$ is calculated by sampling $5000$ sequences and computing their MFE
structures.
The standard error of measuring $\text{IFR}_{\sigma,k}^S$ by sampling $5000$ times is $\pm0.0065$ (derived by computing the
$\text{IFR}_{\sigma,5}^S$ of aae-let-7 sequence-structure pairs $100$ times). By computing the
$\text{IFR}_{\sigma,k}^S$ for each $1\leq k\leq 20$, we obtain the $\text{IFR}_{\sigma,k}^S$-spectrum of the sequence-structure pairs,
$(\sigma,S)$.

\subsection{miRNA data.}
We consider $401$ miRNA precursor sequences of the let-$7$ gene family, obtained from the miRBase database \cite{kozomara2013mirbase}.
These originate from $82$ different animal species. In vertebrate and urochordates multiple homologous miRNA genes are
commonly observed, while single miRNA let-$7$ genes were more common in other animal species. We provide in the supplemental
materials (SM) the numbers of the respective let-$7$ genes (see Table 1 in SM). The secondary structures of these sequences are derived using a
MFE folding algorithm, employing the Turner energy model \cite{Turner:10}.
We compute the $\text{IFR}_{\sigma,k}^S$-spectra, for $1\leq k\leq 20$, for all $401$ let-$7$ sequence-structure pairs.

%%%%
%%%%%%%%%%%%%%%%%%%%%%%%%%%%%%%%%%%%%%%%%%%%%%%%%%%%%%%%%%%%%%%%%%%%%%%%%%%%%%%%%%%%%%%%%%%%%%%%%%%%%%%%%%%%%%%%%%%%%%%%%%%%%
%%%%
\section{RESULTS}
%%%%
%%%%%%%%%%%%%%%%%%%%%%%%%%%%%%%%%%%%%%%%%%%%%%%%%%%%%%%%%%%%%%%%%%%%%%%%%%%%%%%%%%%%%%%%%%%%%%%%%%%%%%%%%%%%%%%%%%%%%%%%%%%%%
%%%%
%3 pages

\subsection{Robustness of native let-7 sequence-structure pairs.}
\begin{figure}[ht]
\begin{center}
\includegraphics[width=70mm]{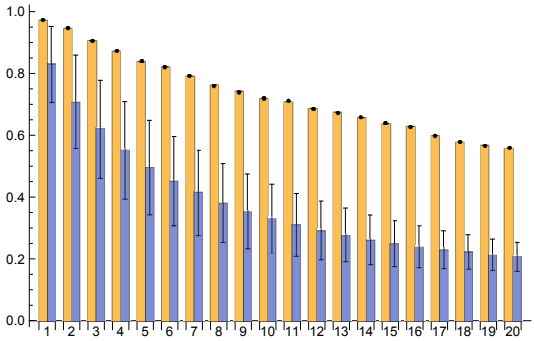}
\end{center}
\caption{The $\text{IFR}_{\sigma,k}^S$-spectrum: the $x$-axis denotes the Hamming distance and the $y$-axis the inverse folding rate.
 The $\text{IFR}_{\sigma,k}^S$-spectrum of the api-let-7 (Acyrthosiphon pisum) gene (yellow) and
  the spectrum of means of the corresponding control set, $\mu(\text{IFR}_{\sigma^*,k}^S)$ (blue).
  The error bar denotes the standard deviation of the control set of $\{\text{IFR}_{\sigma^*,k}^S\}$ for each $k$.
  Since $\text{IFR}_{\sigma,k}^S > \mu(\text{IFR}_{\sigma^*,k}^S)$, api-let-7 is $k$-robust, for all $1\leq k\leq 20$.
  $\text{IFR}_{\sigma,k}^S$ decreases slower than $\mu(\text{IFR}_{\sigma^*,k}^S)$, as $k$ increases.
 } 
\label{F:NVR}
\end{figure}

Borenstein and Ruppin \cite{borenstein2006direct} report that miRNA sequences exhibit a high level of neutrality in comparison
with random sequences folding into a similar structure, suggesting that native miRNA sequence are more robust against single-point mutations.
These results motivate further analysis of native and random sequences, in particular whether or not this neutrality is confined
to a local neighborhood. To this end we uniformly select $50$ of the $401$ native sequence-structure pairs of the let-7 miRNA
family (see Table 2 in the SM).
For each such native pair, $(\sigma,S)$, we derive $100$ random sequences, $\sigma^*$, whose MFE structure is identical to $S$,
as control set. These control sequences are computed using the inverse folding algorithm presented in \cite{barrett2017sequence,garcia2016rnadualpf,levin2012global}.
Given structure $S$, we compute the partition function of all sequences with respect to $S$. We then Boltzmann sample
sequences and filter (by rejection) those sequences, that fold into $S$.
We then compute the $\text{IFR}_{\sigma^*,k}^S$-spectra for the inverse folding solutions and compare for each native
sequence-structure pair $(\sigma,S)$, the $\text{IFR}_{\sigma,k}^S$-spectrum with $\mu(\text{IFR}_{\sigma^*,k}^S)$-spectrum, i.e.~the
mean of all $100$ spectra, $\text{IFR}_{\sigma^*,k}^S$, taken at each respective $k$. We call $(\sigma,S)$ $k$-robust, if
$\text{IFR}_{\sigma,k}^S>\mu(\text{IFR}_{\sigma^*,k}^S)$. We then check the native sequence-structure pairs for $k$-robustness
and quantify the significance of $k$-robustness via $Z$-tests. In Figure \ref{F:NVR} we depict a sequence-structure pair,
that is $k$-robust for $1\leq k\leq 20$. 

%%%
%%%%%%%%%%%%%%%%%%%%%%%%%%%%%%%%%%%%%%%%%%%%%%%%%%%%%%%%%%%%%%%%%%%%%%%%%%%%%%%%%%%%%%%%%%%%%%%%%%%%%%%%%%%%%%%%%%%%%%%%%%%%%%%
%%%
%%%
%%%%%%%%%%%%%%%%%%%%%%%%%%%%%%%%%%%%%%%%%%%%%%%%%%%%%%%%%%%%%%%%%%%%%%%%%%%%%%%%%%%%%%%%%%%%%%%%%%%%%%%%%%%%%%%%%%%%%%%%%%%%%%%
%%%

In Table \ref{table:01} we summarize the data on $k$-robustness of the $50$ let-7 sequence-structure pairs
included in our study. We observe that $96\%$ of the selected let-7 miRNAs are $1$-robust. This observation is consistent
with \cite{borenstein2006direct}, providing further evidence that native let-7 sequence-structure pairs are robust
against single-point mutations.
However, the mutational robustness of native let-7 sequence-structure pairs is not restricted to single-point mutations:
for all $1\leq k\leq 20$, we observe that $k$-robustness holds for over $90\%$ of the selected sequence-structure pairs.
This suggests, that the mutational robustness of let-7 sequence-structure pairs is not a local phenomenon.
Furthermore, the percentage of significantly $k$-robust genes increases as $k$ increase, see Table \ref{table:01}.
In Figure \ref{F:NVR} we display the $\text{IFR}^S_{\sigma,k}$-spectrum of the api-let-7 sequence-structure pair, illustrating
the aforementioned phenomenon. The api-let-7 sequence-structure pair is $1$-robust but not significantly $1$-robust. However,
for $5\leq k \leq 20$, the api-let-7 sequence-structure pairs is significantly $k$-robust.

%%%
%%%%%%%%%%%%%%%%%%%%%%%%%%%%%%%%%%%%%%%%%%%%%%%%%%%%%%%%%%%%%%%%%%%%%%%%%%%%%%%%%%%%%%%%%%%%%%%%%%%%%%%%%%%%%%%%
%%%%

\begin{table}[b]
\tableparts{%
\caption{$k$-robustness of native sequence-structure pairs}
\label{table:01}%
}{%
\begin{tabular*}{\columnwidth}{@{}lllllllllllll@{}}
\toprule
Hamming distance \hspace{0.8cm} &  $k$-robust \hspace{1cm} & $p<0.05$ %& $p<0.01$ 
\\
%& (\%) & (s$^{-1}$) 
%\\
\colrule
k=1 & 96\% & 8\% %& 0\% 
\\
k=2 & 96\% & 28\% %& 2\% 
\\
k=3 & 94\% & 38\% %& 10\%
\\
k=4 & 92\% & 38\% %& 16\%
\\
k=5 & 94\% & 38\% %& 20\% 
\\
k=10 & 92\% & 48\% %& 28\% 
\\
k=15 & 94\% & 56\% %& 38\% 
\\
k=20 & 96\% & 58\% %& 54\% 
\\
\botrule
\end{tabular*}%
}
{Second column: the percentage of $k$-robust ($\text{IFR}_{\sigma,k}^S > \mu(\text{IFR}_{\sigma^*,k}^S))$ genes.
 Third column: the percentage of the significantly $k$-robust genes for $p<0.05$. The $p$ values denote
 the probability of observing $\text{IFR}_{\sigma^*,k}^S > \text{IFR}_{\sigma,k}^S$ by chance and are calculated via $Z$-tests.}
\end{table}
%%%
%%%%%%%%%%%%%%%%%%%%%%%%%%%%%%%%%%%%%%%%%%%%%%%%%%%%%%%%%%%%%%%%%%%%%%%%%%%%%%%%%%%%%%%%%%%%%%%%%%%%%%%%%%%%%%%%
%%%%

\begin{figure*}[ht]
\begin{center}
  \begin{tabular}{ccc}

    \includegraphics[width=0.65\columnwidth]{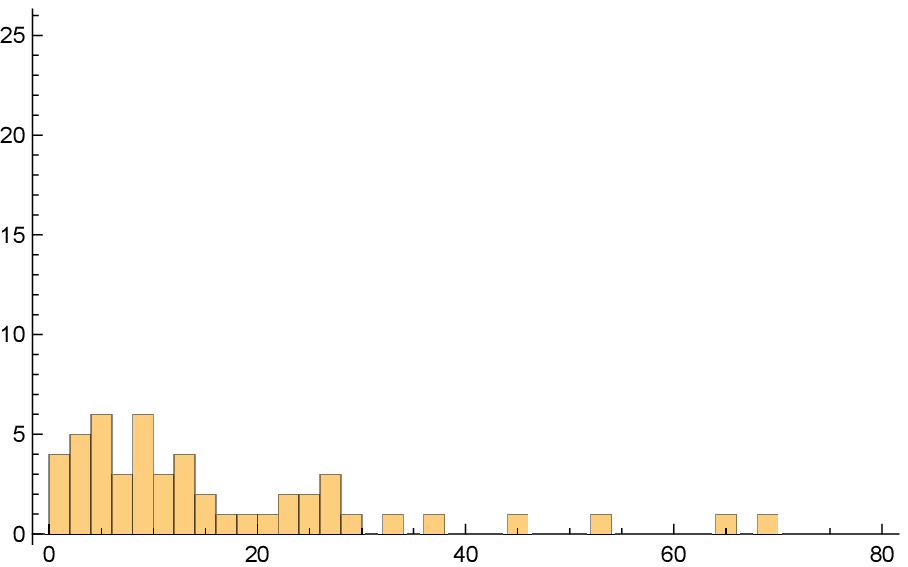}&

    \includegraphics[width=0.65\columnwidth]{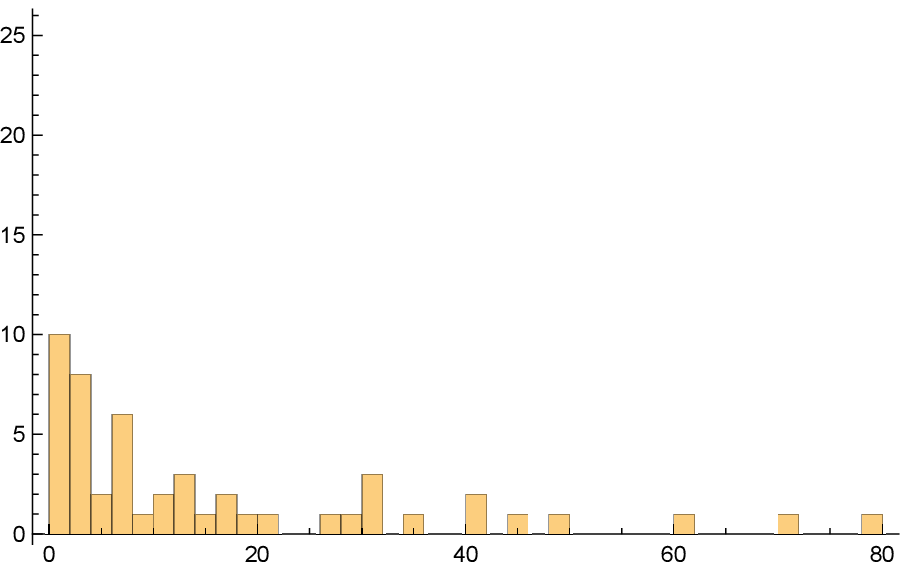}&

    \includegraphics[width=0.65\columnwidth]{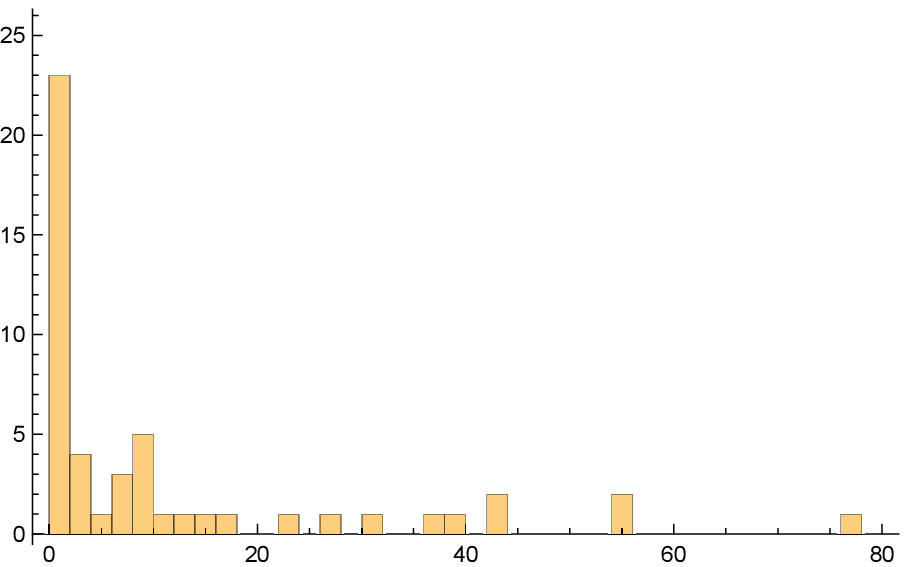}\\
        $k=1$ &
$k=5$ &
$k=20$\\

  \end{tabular}
  \end{center}
  \caption{
The distribution of $\text{IFR}_{\sigma,k}^S$-ranks, $r_k$, of the native sequence-structure pairs $\text{IFR}_{\sigma,k}^S$.
The $x$-axis denotes ranking and the $y$-axis frequency, $k=1$ (Left), $k=5$ (Center) and $k=20$ (Right). 
For each native sequence-structure pair, $(\sigma,S)$, the $\text{IFR}_{\sigma,k}^S$ is ranked among $100$
$\text{IFR}_{\sigma^*,k}^S$-values, where $\sigma^*$ is a random sequence whose MFE structure is $S$. 
}
\label{F:rank}
\end{figure*}

We proceed by considering for each native sequence-structure pair $(\sigma,S)$, $r_k$, its $\text{IFR}_{\sigma,k}^S$-rank
among the $\text{IFR}_{\sigma^*,k}^S$-values, respectively. Figure \ref{F:rank} presents the
distribution of $r_k$, for $k=1,5,20$.
Examination of the rank distribution shows that native pairs exhibit a propensity towards high ranks for each $k$,
supporting the observation of $k$-robustness. As $k$ increases, the propensity towards high ranks becomes more and
more pronounced. In case of $k=1$, only $4$ out of $50$ native sequence-structure pairs have $r_1=1$. For $k=20$, however,
this quantity increases to $23$. This is consistent with the increasing percentage of significantly $k$-robust
genes as $k$ increases, further demonstrating that native let-7 sequence-structure pairs exhibit mutational robustness
against multiple-point mutations.

%%%
%%%%%%%%%%%%%%%%%%%%%%%%%%%%%%%%%%%%%%%%%%%%%%%%%%%%%%%%%%%%%%%%%%%%%%%%%%%%%%%%%%%%%%%%%%%%%%%%%%%%%%%%%%%%%%%%
%%%%

%%%
%%%%%%%%%%%%%%%%%%%%%%%%%%%%%%%%%%%%%%%%%%%%%%%%%%%%%%%%%%%%%%%%%%%%%%%%%%%%%%%%%%%%%%%%%%%%%%%%%%%%%%%%%%%%%%%%
%%%%

By comparing specific native sequence-structure pairs with the respective control pairs (obtained by inverse folding),
we observe native sequence-structure pairs exhibit in general higher $\text{IFR}_{\sigma,k}^S$-values.
In this analysis the sequence-structure pair, $(\sigma,S)$, is fixed and we contrast native pairs with
pairs obtained via random inverse folded sequences.

\begin{figure*}[ht]
\begin{center}
  \begin{tabular}{ccc}

    \includegraphics[width=0.65\columnwidth]{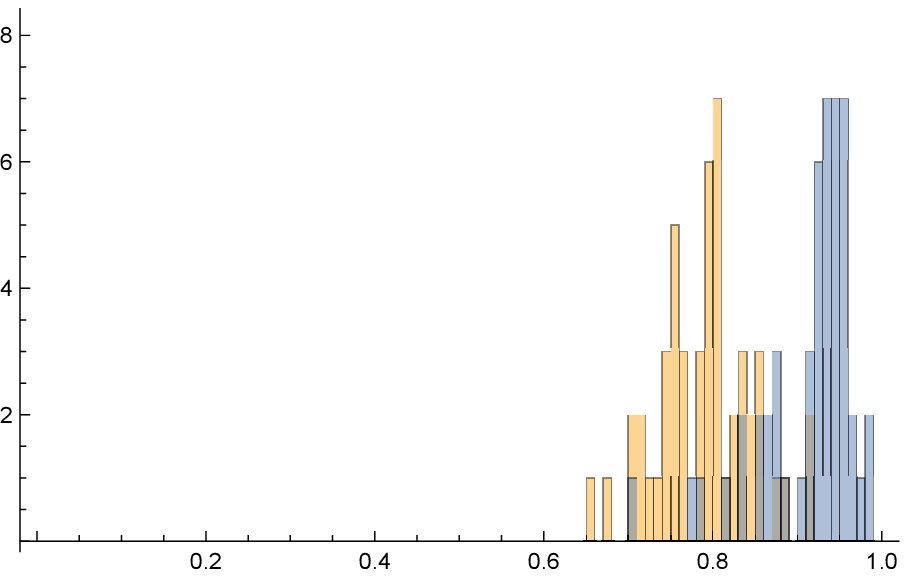}&

    \includegraphics[width=0.65\columnwidth]{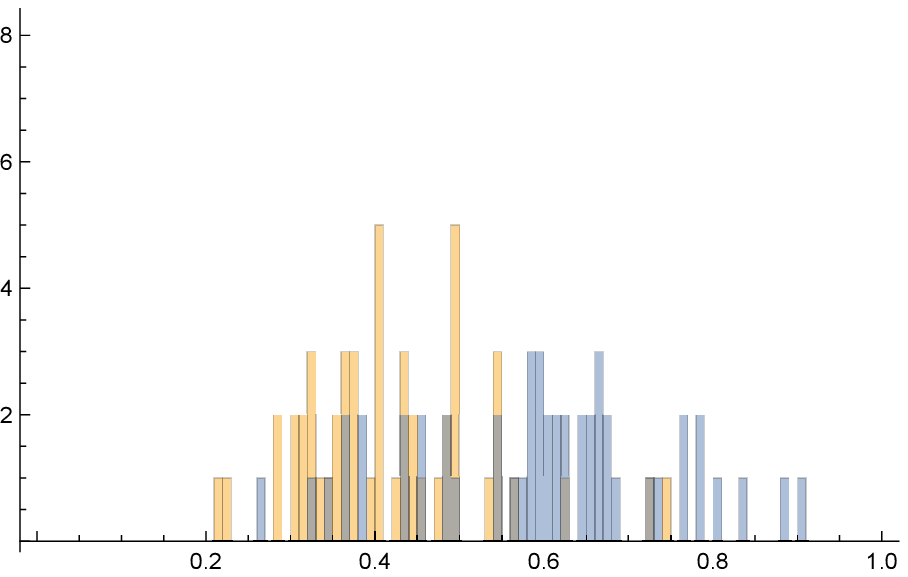}&

    \includegraphics[width=0.65\columnwidth]{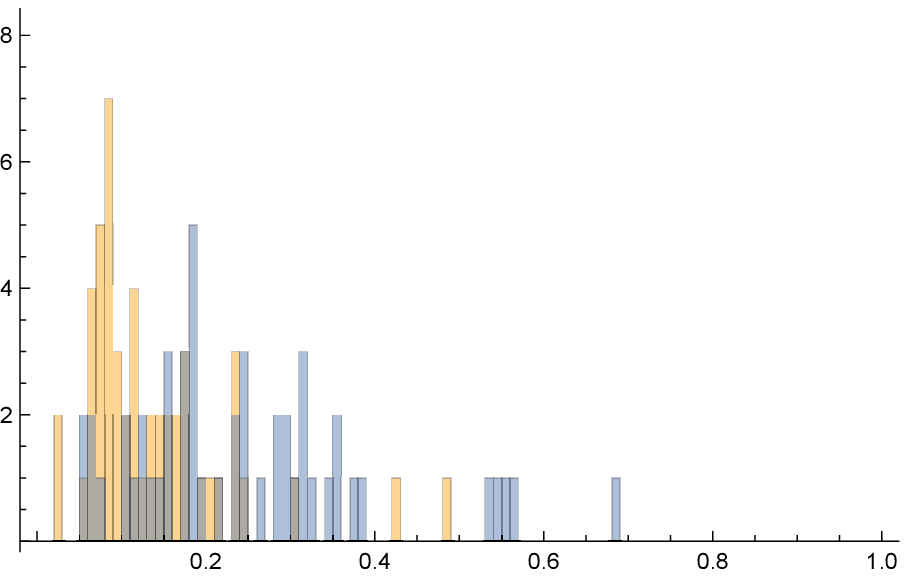}\\
    $k=1$ &
$k=5$ &
$k=20$\\

  \end{tabular}
  \end{center}
  \caption{
  The distribution of $\text{IFR}_{\sigma,k}^S$, of native pairs (blue) and $\mu(\text{IFR}_{\sigma^*,k}^S)$, of the control
  sets (yellow). The $x$-axis denotes the inverse folding rate and the $y$-axis denotes the frequency,
  $k=1$ (Left) $k=5$ (Center) $k=20$ (Right). 
}
\label{Histogram}
\end{figure*}

We can augment this analysis by considering the {\it ensemble} of spectra of native pairs versus the {\it ensemble} of all
inverse folded sequence-structure pairs. In that, for any fixed $k$, we can integrate the information of all native pairs
contrasting this with the integrated information of the inverse folded pairs. Figure \ref{Histogram} displays these two
distributions: $\text{IFR}_{\sigma,k}^S$ of native pairs and $\mu(\text{IFR}_{\sigma^*,k}^S)$, of the inverse folded pairs
for $k=1,5,20$. For each $k$, we not only observe that the mean of the $\text{IFR}_{\sigma,k}^S$ is greater than
  that of the terms $\mu(\text{IFR}_{\sigma^*,k}^S)$, but also that the distributions of $\text{IFR}_{\sigma,k}^S$ and
$\mu(\text{IFR}_{\sigma^*,k}^S)$ are distinctively different. Furthermore, the difference between the two distributions
for each $k$ is statistically significant, see Table \ref{table:distribution}, where the $p$ value is calculated by
two tailed Wilcoxon signed rank tests for paired data. 
%%%
%%%%%%%%%%%%%%%%%%%%%%%%%%%%%%%%%%%%%%%%%%%%%%%%%%%%%%%%%%%%%%%%%%%%%%%%%%%%%%%%%%%%%%%%%%%%%%%%%%%%%%%%%%%%%%%%
%%%%

%%%
%%%%%%%%%%%%%%%%%%%%%%%%%%%%%%%%%%%%%%%%%%%%%%%%%%%%%%%%%%%%%%%%%%%%%%%%%%%%%%%%%%%%%%%%%%%%%%%%%%%%%%%%%%%%%%%%
%%%%
%%%
%%%%%%%%%%%%%%%%%%%%%%%%%%%%%%%%%%%%%%%%%%%%%%%%%%%%%%%%%%%%%%%%%%%%%%%%%%%%%%%%%%%%%%%%%%%%%%%%%%%%%%%%%%%%%%%%
%%%%	
\begin{table}[b]
\tableparts{%
\caption{Distinct distribution of $\text{IFR}_{\sigma,k}^S$ and $\mu(\text{IFR}_{\sigma^*,k}^S)$}
\label{table:distribution}%
}{%
\begin{tabular*}{\columnwidth}{@{}lllllllllllll@{}}
\toprule
Hamming distance & mean $\text{IFR}_{\sigma,k}^S$ & mean $\mu(\text{IFR}_{\sigma^*,k}^S)$ & $p$ value 
\\
%& (\%) & (s$^{-1}$) 
%\\
\colrule
k=1 & 0.9118 & 0.7904 &  1.09$\times 10^{-9}$
\\
k=5 & 0.5952 & 0.4220 & 3.57$\times 10^{-9}$
\\
k=20 & 0.2483 & 0.1387 & 2.23$\times 10^{-9}$
\\
\botrule
\end{tabular*}%
}
{The second and third columns display the mean value of $\text{IFR}_{\sigma,k}^S$ and $\mu(\text{IFR}_{\sigma^*,k}^S)$,
  respectively, for each $k$. $p$ values denote the probability of observing more extreme differences between
  $\{\text{IFR}_{\sigma,k}^S\}$ and $\{\mu(\text{IFR}_{\sigma^*,k}^S)\}$ at random, assuming two samples are drawn
  from the same distribution. The $p$ value is calculated by two tailed Wilcoxon signed rank test for paired data.
}
\end{table}
%%%
%%%%%%%%%%%%%%%%%%%%%%%%%%%%%%%%%%%%%%%%%%%%%%%%%%%%%%%%%%%%%%%%%%%%%%%%%%%%%%%%%%%%%%%%%%%%%%%%%%%%%%%%%%%%%%%%
%%%%

The increase of significantly $k$-robust native sequence-structure pairs for increasing $k$, as well as the tendency of native $\text{IFR}_{\sigma,k}^S$,
to assume high ranks, for increasing $k$, suggest that $k$-robustness, $2\leq k\leq 20$, is not a byproduct of
$1$-robustness.

Clearly, the very notion of $\text{IFR}_{\sigma,k}^S$-spectrum raises the question to what extend
$k$-robustness of native sequence-structure pairs, $2\leq k\leq 20$, is strongly correlated to $1$-robustness.
In other words, to what extent is robustness against multiple-point mutations induced by robustness against single-point mutations.

In order to quantify this, we conduct a systematic correlation analysis for $\text{IFR}_{\sigma,k}^S$ and
$\mu(\text{IFR}_{\sigma^*,k}^S)$. Regarding $\mu(\text{IFR}_{\sigma^*,k}^S)$ as the intrinsic $k$-robustness
of the structure $S$, we arrive at interpreting the term $\text{IFR}_{\sigma,k}^S-\mu(\text{IFR}_{\sigma^*,k}^S)$ as
adaptive $k$-robustness. 

The intrinsic structural $k$-robustness $\mu(\text{IFR}_{\sigma^*,k}^S)$ between different $k$ exhibits very strong
correlation. Point in case: Spearman's rank correlation coefficient between $\mu(\text{IFR}_{\sigma^*,1}^S)$ and
$\mu(\text{IFR}_{\sigma^*,20}^S)$ is $0.8426$, where $p<10^{-13}$.
However, the correlation between $\text{IFR}_{\sigma,1}^S-\mu(\text{IFR}_{\sigma^*,1}^S)$ and
$\text{IFR}_{\sigma,k}^S-\mu(\text{IFR}_{\sigma^*,k}^S)$, drops much faster as $k$ increases. The Spearman's rank correlation
coefficient between $\text{IFR}_{\sigma,1}^S-\mu(\text{IFR}_{\sigma^*,1}^S)$ and $\text{IFR}_{\sigma,20}^S-
\mu(\text{IFR}_{\sigma^*,20}^S)$ is $0.3693$; where $p=0.0083$, showing weak correlation.
This indicates that there exists factors beyond $1$-robustness that contribute to the $k$-robustness of {\it native}
sequence-structure pairs.

We finally study the role of the free energy for $k$-robustness. \cite{bonnet2004evidence} reports that miRNA exhibits increased
thermodynamic stability and furthermore that mutations in the HAR further stabilize the structure \cite{pollard2006forces}. This gives rise to
the question whether increased $k$-robustness is a result of the increased thermodynamic stability of miRNAs.

\begin{figure}[ht]
\begin{center}
\includegraphics[width=1.0\columnwidth]{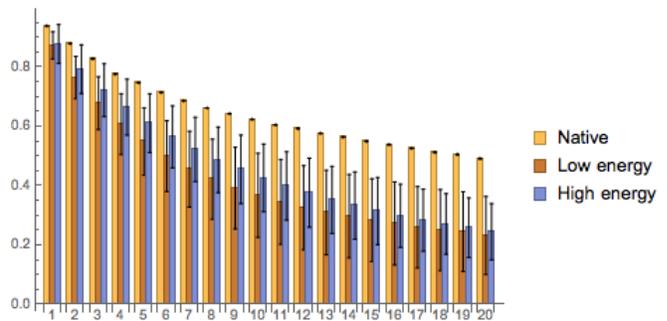}
\end{center}
\caption{Spectra of a native sequence-structure pair versus those of inverse folding solutions having lower and higher
  energies. We display the spectrum of the native, let-7 miRNA pair (yellow), the spectrum of the mean $\text{IFR}_{\sigma,k}^S$ of inverse
  fold solutions having lower (red) and higher (blue) energy, respectively and their standard deviations. The $x$-axis
  denotes the Hamming distance and the $y$-axis the inverse folding rate. 
}
\label{F:IFRnative}
\end{figure}

To test
this, we select a native sequence-structure pair from the miRNA let-7 family and generate two sets of
sequences: $\Sigma_{\text{high}}$ and $\Sigma_{\text{low}}$, each consisting of $30$ sequences that fold into $S$, having
higher and lower energy than the native pair, respectively.
It turns out, that sequences generated by RNAinverse \cite{lorenz2011viennarna} tend to have higher energy than the native pair,
while sequences generated by the dual Boltzmann sampler, filtered by rejection to fold into $S$, tend to have lower
energy. We compute the mean and the standard deviation of the spectra of the sequences of $\Sigma_{\text{low}}$
and $\Sigma_{\text{high}}$ respectively and integrate our findings in Figure \ref{F:IFRnative}. 

%%%
%%%%%%%%%%%%%%%%%%%%%%%%%%%%%%%%%%%%%%%%%%%%%%%%%%%%%%%%%%%%%%%%%%%%%%%%%%%%%%%%%%%%%%%%%%%%%%%%%%%%%%%%%%%%%%%%
%%%%
%%%
%%%%%%%%%%%%%%%%%%%%%%%%%%%%%%%%%%%%%%%%%%%%%%%%%%%%%%%%%%%%%%%%%%%%%%%%%%%%%%%%%%%%%%%%%%%%%%%%%%%%%%%%%%%%%%%%
%%%%
Figure \ref{F:IFRnative} shows that the $\text{IFR}_{\sigma,k}^S$-spectrum of the native pair is distinctively higher than the ones derived from
$\Sigma_{\text{low}}$ and $\Sigma_{\text{high}}$. We have confirmed this result for $10$ additional native sequence-structure
pairs. The findings suggest that thermodynamic stability is not the sole factor, native pairs have been selected
for. The mutational robustness of native pairs is not a byproduct of evolving toward thermodynamic stability.

%%%
%%%%%%%%%%%%%%%%%%%%%%%%%%%%%%%%%%%%%%%%%%%%%%%%%%%%%%%%%%%%%%%%%%%%%%%%%%%%%%%%%%%%%%%%%%%%%%%%%%%%%%%%%%%%%%%%
%%%%
\subsection{Robustness of metazoan species.}
%%%
%%%%%%%%%%%%%%%%%%%%%%%%%%%%%%%%%%%%%%%%%%%%%%%%%%%%%%%%%%%%%%%%%%%%%%%%%%%%%%%%%%%%%%%%%%%%%%%%%%%%%%%%%%%%%%%%
%%%%

\cite{freilich2010decoupling} studies genetic robustness of networks of bacterial genes, observing variations among species
in their level of genetic robustness, reflecting adaptations to different ecological niches and lifestyles. The genetic
robustness of a network refers to its ability to buffer mutations via the existence of alternative pathways.
The species-specific variations raise the question whether such variations can also be observed for the $\text{IFR}_{\sigma,k}^S$-spectra
across different animal species. Are $\text{IFR}_{\sigma,k}^S$-spectra to some extent a reflection of phylogenetic relationships?

\begin{figure}[ht]
\begin{center}
\includegraphics[width=1.0\columnwidth]{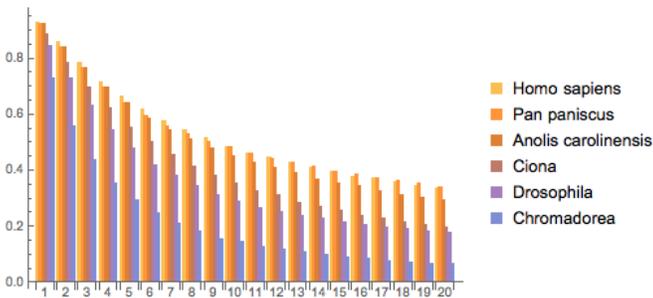}
\end{center}
\caption{Mean $\text{IFR}_{\sigma,k}^S$ within each of the six taxa, from left to right:
  Homo sapiens, Pan paniscus, Anolis carolinensis, Ciona, Drosophila, Chromadorae.
  The $x$-axis denotes the Hamming distance and the $y$-axis the inverse folding rate.
  The relative ordering of $\text{IFR}_{\sigma,k}^S$-values among taxa is consistent for each $k$ (from high to low: Homo sapiens, Pan paniscus,
  Anolis carolinensis, Ciona, Drosophila, Chromadorae, the only exception being $\text{IFR}_{\sigma,k}^S$-values of Pan paniscus becoming slightly
  larger than those of Homo sapiens, for $14\leq k \leq 20$). 
} 
\label{F:cross_spe}
\end{figure}

\begin{figure*}[ht]
\begin{center}
  \begin{tabular}{ccc}

    \includegraphics[width=0.66\columnwidth]{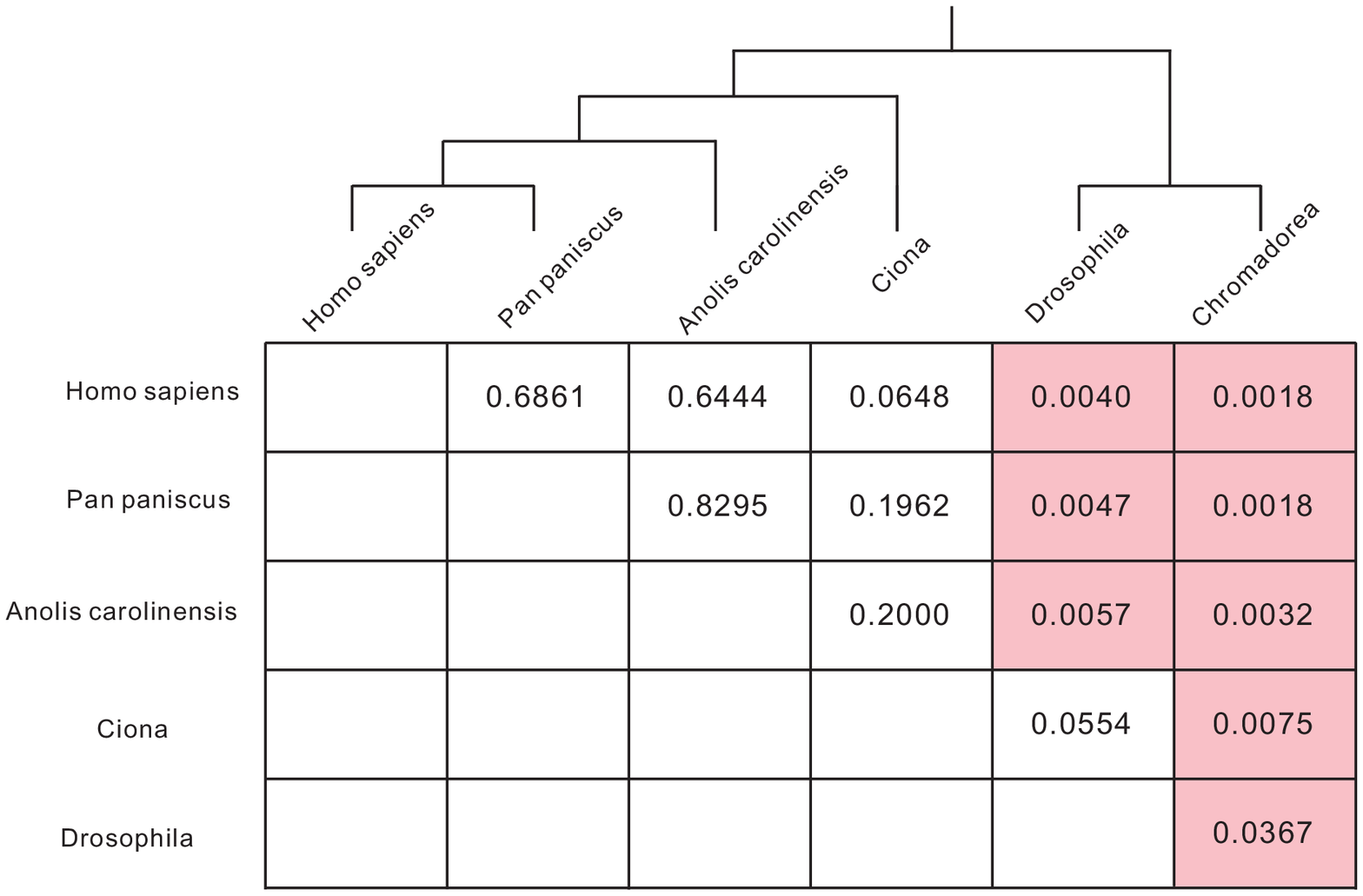}&

    \includegraphics[width=0.66\columnwidth]{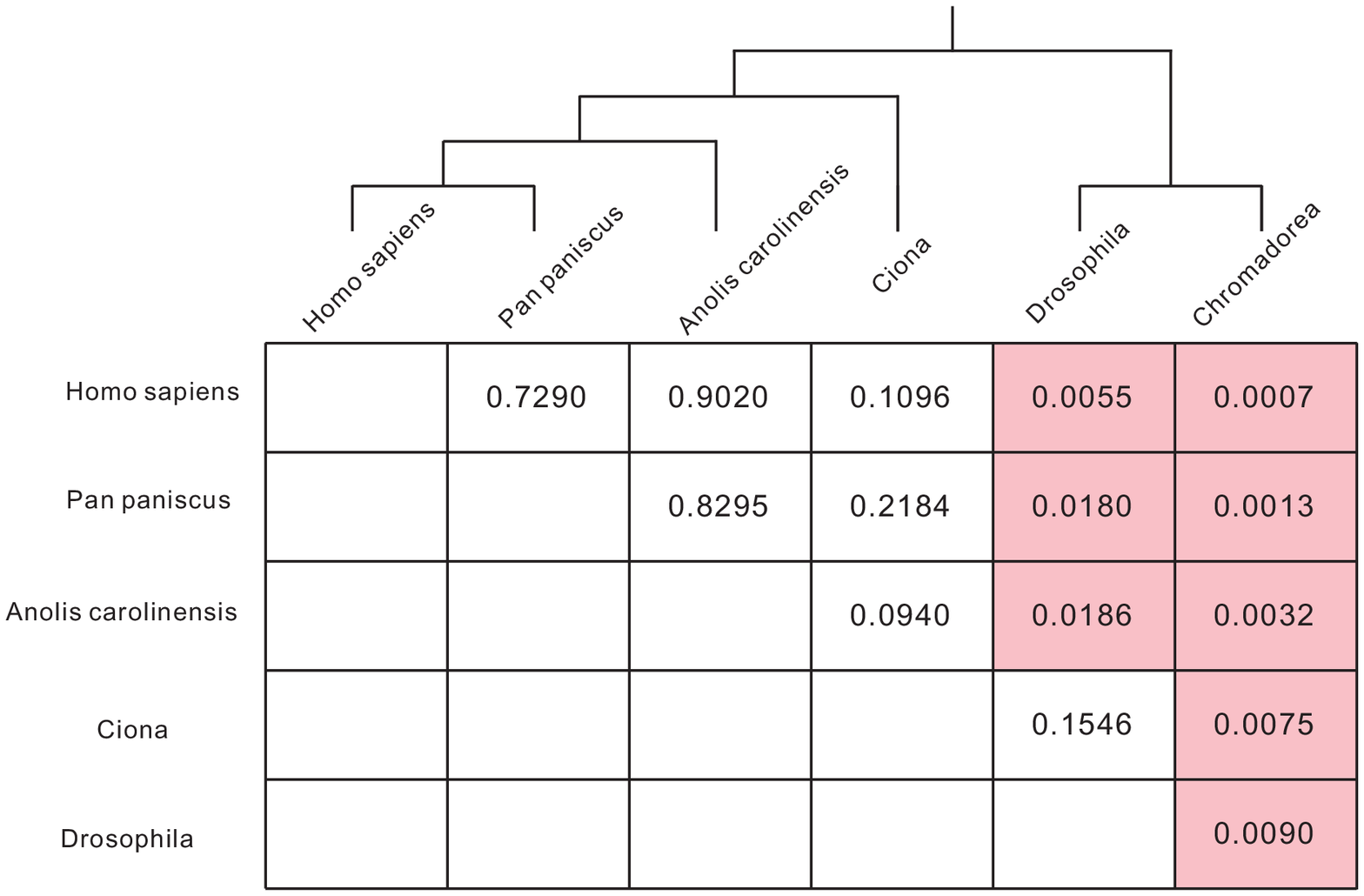}&

    \includegraphics[width=0.66\columnwidth]{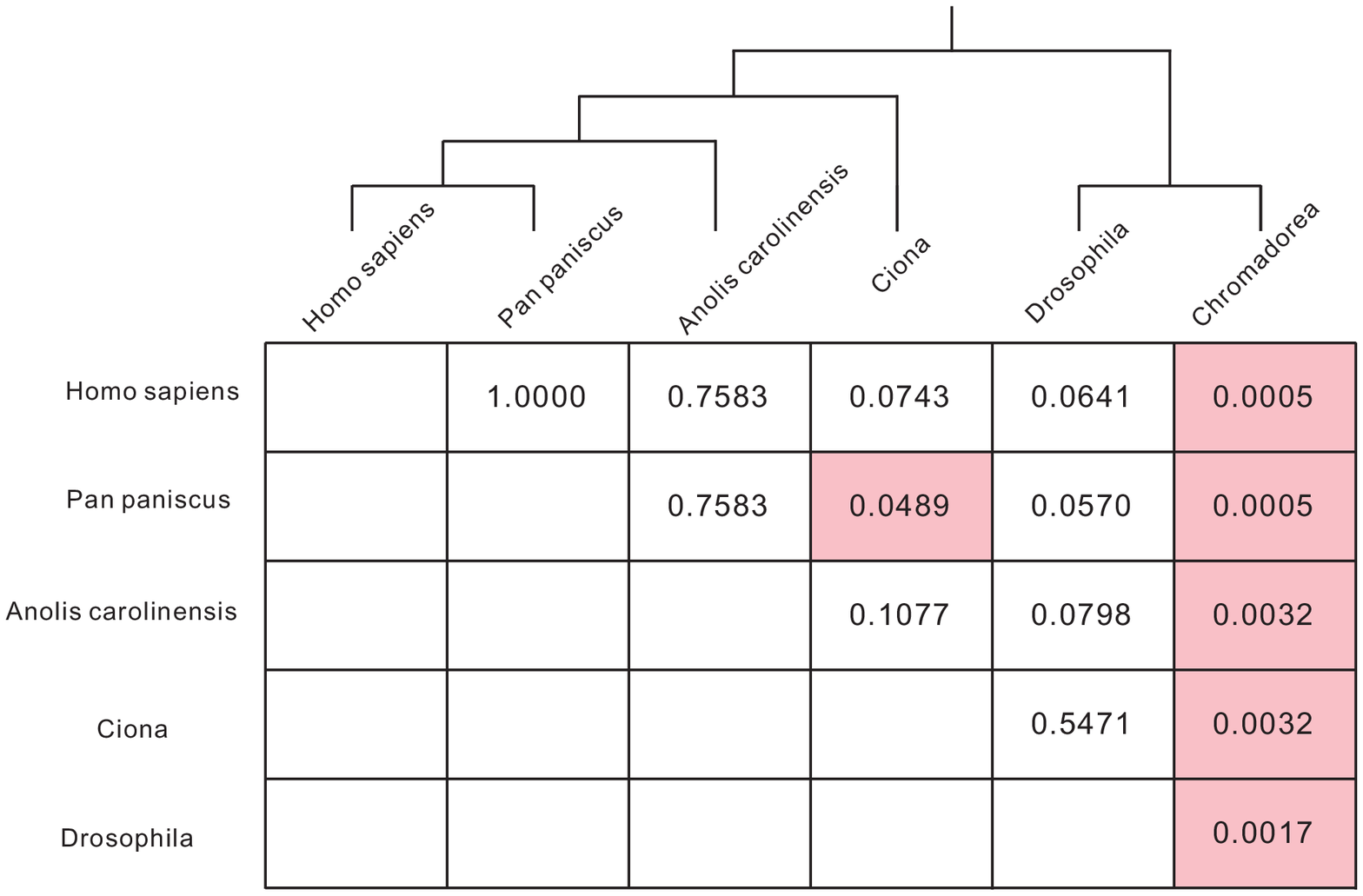}\\
        $k=1$ &
$k=5$ &
$k=20$\\

  \end{tabular}
  
  \end{center}
  \caption{Statistical significance ($p$ value) of the difference within the let-7 genes $\text{IFR}_{\sigma,k}^S$-distribution
  between taxa, for $k=1$ (left), $k=5$ (middle) and $k=20$ (right). $p$ values denote the probability of observing a larger
  difference in $\text{IFR}_{\sigma,k}^S$-distribution between two corresponding taxa at random, assuming they are drawn from
  the same probability distribution, computed by two tailed Wilcoxon rank sum test. Differences between two taxa, that are
  statistically significant $(p<0.05)$, are highlighted and phylogenetic relations among the six taxa are displayed.
}
\label{F: taxa_test}
\end{figure*}

Herein, we perform an evolutionary analysis of the $\text{IFR}_{\sigma,k}^S$-spectra of let-7 miRNA sequence-structure pairs across different
animal species. We first perform our analysis on $6$ selected taxa: Homo sapiens, Pan paniscus, Anolis carolinensis, Ciona,
Drosophila, Chromadorae. There are $12$, $12$, $11$, $11$, $14$ and $7$ let-7 genes found in the miRBase within each
taxon, respectively.
For almost all $1\leq k\leq 20$, we observe the mean $\text{IFR}_{\sigma,k}^S$-value within each taxa to decrease from Homo sapiens to
Chromadorae, see Figure \ref{F:cross_spe} and Table.~\ref{table:03}. The only exceptions are the $\text{IFR}_{\sigma,k}^S$-values of
Pan paniscus, that become slightly larger than those of Homo sapiens, for $14\leq k \leq 20$.
We observe that ``higher'' metazoan species, as
Homo sapiens, Pan paniscus, Anolis carolinensis, Ciona, all of which being Chordata, exhibit larger $\text{IFR}_{\sigma,k}^S$-values than ``lower''
metazoan species, as Drosophila, Chromadorae, all of which being Ecdysozoa.
%%%
%%%%%%%%%%%%%%%%%%%%%%%%%%%%%%%%%%%%%%%%%%%%%%%%%%%%%%%%%%%%%%%%%%%%%%%%%%%%%%%%%%%%%%%%%%%%%%%%%%%%%%%%%%%%%%%%
%%%%
%%%
%%%%%%%%%%%%%%%%%%%%%%%%%%%%%%%%%%%%%%%%%%%%%%%%%%%%%%%%%%%%%%%%%%%%%%%%%%%%%%%%%%%%%%%%%%%%%%%%%%%%%%%%%%%%%%%%
%%%%
%%%
%%%%%%%%%%%%%%%%%%%%%%%%%%%%%%%%%%%%%%%%%%%%%%%%%%%%%%%%%%%%%%%%%%%%%%%%%%%%%%%%%%%%%%%%%%%%%%%%%%%%%%%%%%%%%%%%
%%%%
\begin{table}[b]
\tableparts{%
\caption{Mean $\text{IFR}_{\sigma,k}^S$ within each taxa at each $k$}
\label{table:03}%
}{%
\begin{tabular*}{\columnwidth}{@{}lllllllllllll@{}}
\toprule
 & $k=1$ & $k=5$ & k=20 
\\
%& (\%) & (s$^{-1}$) 
%\\
\colrule
Homo sapiens & 0.9288 & 0.6632 &  0.3356
\\
Pan paniscus & 0.9234 & 0.6404 &  0.3402
\\
Anolis carolinensis & 0.9211 & 0.6412 &  0.2949
\\
Ciona & 0.8874 & 0.5509 &  0.1971
\\
Drosophila & 0.8524 & 0.4889 &  0.1802
\\
Chromadorae & 0.9118 & 0.7904 &  0.0642
\\
\botrule
\end{tabular*}%
}
{The table displays the mean value of $\text{IFR}_{\sigma,k}^S$ of the let-7 sequence-structure pairs within each taxa,
 at $k=1,5,20$.}
\end{table}
%%%
%%%%%%%%%%%%%%%%%%%%%%%%%%%%%%%%%%%%%%%%%%%%%%%%%%%%%%%%%%%%%%%%%%%%%%%%%%%%%%%%%%%%%%%%%%%%%%%%%%%%%%%%%%%%%%%%
%%%%
In addition, the difference of $\text{IFR}_{\sigma,k}^S$-values among evolutionary closely related species, as Homo sapiens, Pan paniscus and
Anolis carolinensis are small. However, comparing taxa, less related in the phylogenetic tree of life, the difference
becomes significant. Statistical tests (two tailed Wilcoxon rank sum test) are conducted to quantify the statistical
significance of this difference within the let-7 genes $\text{IFR}_{\sigma,k}^S$-value distribution among these six taxa, see
Figure \ref{F: taxa_test}, for $k=1,5,20$.

The results of statistical tests at $k=1,5,20$ demonstrate that $\text{IFR}_{\sigma,k}^S$-distributions of the six taxa
reflect the complexity of the organisms and their phylogenetic relations. Homo sapiens, Pan paniscus and Anolis
carolinensis almost always exhibit significantly higher $\text{IFR}_{\sigma,k}^S$-values than Drosophila and Chromadorae,
for $k=1,5,20$. On the other hand, differences among Homo sapiens, Pan paniscus and Anolis carolinensis are statistically
insignificant for $k=1,5,20$.
However, we observe, for $k=20$ insignificant differences between Homo sapiens, Pan paniscus and Anolis carolinensis
  with Drosophila, though the $p$ values are very close to $0.05$. A more distinguished variation is observed comparing Ciona
  with Drosophila. In case of $k=1$, the difference is close of being significant with $p=0.0554$, but for $k=20$, the
  difference is insignificant with $p=0.5417$.

\begin{figure*}[ht]
\begin{center}
\includegraphics[width=2\columnwidth]{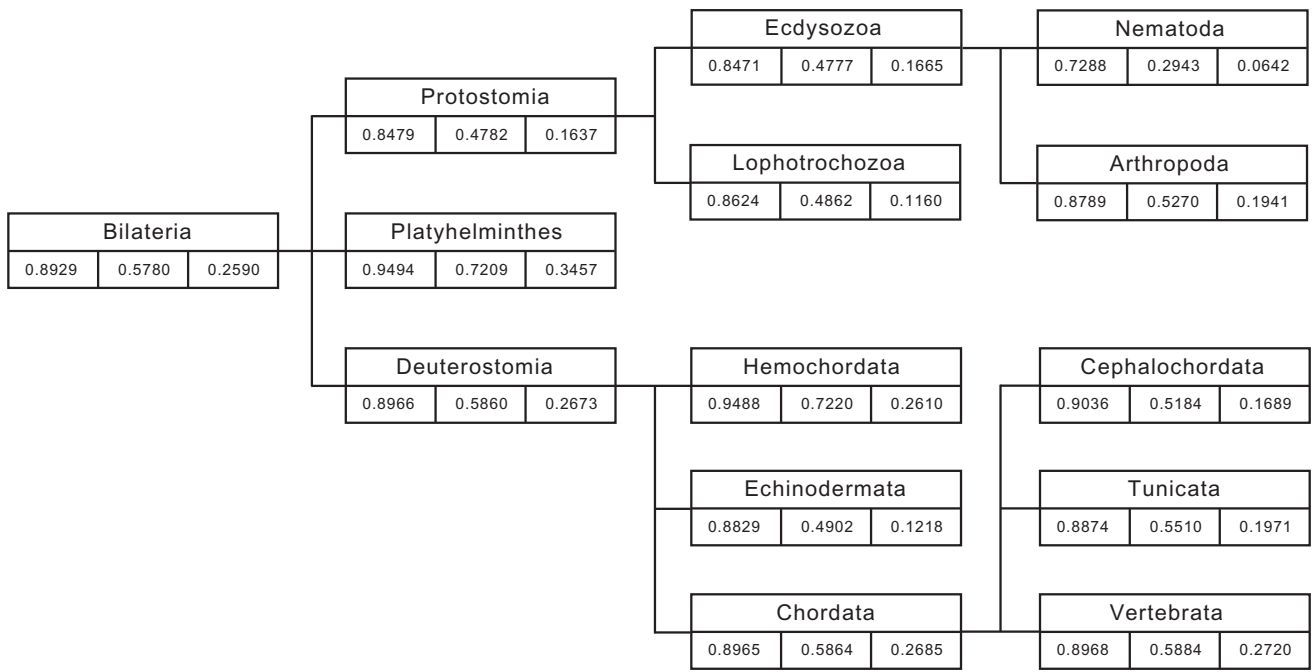}
\end{center}
\caption{Phylogenetic tree of the major taxa and mean $\text{IFR}_{\sigma,k}^S$ of the let-7 sequence-structure pairs
  within each taxon, for $k=1,5,20$. Mean $\text{IFR}_{\sigma,k}^S$-value for $k=1$ (Right hand side), $k=5$ (Center)
  and $k=20$ (Left hand side).
}
\label{F:subtree}
\end{figure*}

We proceed by conducting the above analysis for all $401$ let-7 sequence-structure pairs of the miRBase. Let-7 genes have
been found in $82$ metazoan species. Multiple homologous miRNA genes are commonly observed in vertebrates, while single
miRNA let-7 genes were more common in other animal species. The phylogenetic tree of all $82$ species is constructed
using the ``Interactive Tree Of Life'' (iTOL) \cite{letunic2006interactive,letunic2016interactive}, based on NCBI
Taxonomy \cite{federhen2011ncbi}. In addition, iTOL determines taxonomic classes of all internal nodes. Given the
phylogenetic tree, we not only compute the mean $\text{IFR}_{\sigma,k}^S$-values of the let-7 sequence-structure pairs within
each specific species, but also the mean $\text{IFR}_{\sigma,k}^S$-values within taxa, corresponding to the internal
nodes of the phylogenetic tree. Figure \ref{F:subtree} depicts a subtree of the phylogenetic tree, that includes the major
taxa as well as the mean $\text{IFR}_{\sigma,k}^S$-values within each taxon, for $k=1,5,20$. The complete phylogenetic
tree is presented in the SM (Figure 3,4,5).

For let-7 miRNAs, we observe that higher metazoan species typically exhibit higher $\text{IFR}_{\sigma,k}^S$-values as well as
significant differences in the $\text{IFR}_{\sigma,k}^S$-distribution across taxa.
For instance, $\text{IFR}_{\sigma,k}^S$-distributions of Deuterostomia
  and Protostomia are found to be significantly different ($p=0.00016, 0.00058, 0.00033$ for $k=1,5,20$, calculated by
  two tailed Wilcoxon rank sum tests).

%%%
%%%%%%%%%%%%%%%%%%%%%%%%%%%%%%%%%%%%%%%%%%%%%%%%%%%%%%%%%%%%%%%%%%%%%%%%%%%%%%%%%%%%%%%%%%%%%%%%%%%%%%%%%%%%%%%
%%%
\section{DISCUSSION}
%%%
%%%%%%%%%%%%%%%%%%%%%%%%%%%%%%%%%%%%%%%%%%%%%%%%%%%%%%%%%%%%%%%%%%%%%%%%%%%%%%%%%%%%%%%%%%%%%%%%%%%%%%%%%%%%%%%
%%%

In this paper we augment the analysis of genetic sequences by incorporating structural information.
The information represented by structures is distinctively different from that represented by sequences. 
Structures encode relations between pairs of {\it loci}. Such a relation can be realized in multiple ways, i.e., 
by the bond between loci $i$ and $j$ by AU, UA, GU, UG, CG and GC. In that, one can expect a meaningful enhancement of the
sequence information.

Incorporating structural information, i.e.~considering sequences and structures combined, provides new
ways to analyze genetic material. We investigate mutational robustness of let-7 miRNA, making use of this 
perspective and introduce the $\text{IFR}_{\sigma,k}^S$-spectrum, a novel observable, that quantifies mutational robustness. The
$\text{IFR}_{\sigma,k}^S$-spectrum depends on both: the reference sequence and structure, respectively and allows us to delocalize
the study of mutational robustness beyond single-point mutations.

Our study provides evidence for direct evolution of increased robustness in let-7 miRNAs, by comparing native
let-7 miRNA sequence-structure pairs with the control pairs obtained by inverse folding algorithms. It provides
evidence that mutational robustness is not local, i.e.~it cannot be deduced from or restricted to single-point mutations.
On the contrary, robustness effects become more pronounced in higher Hamming distances:
the percentage of significantly $k$-robust, native let-7 sequence-structure pairs increases as the Hamming
  distance $k$ increases. By conducting a correlation analysis between $\text{IFR}_{\sigma,k}^S$-values, we provide evidence that there exist additional factors
that contribute to mutational robustness against multiple-point mutations. The spectrum itself contains information
that cannot be inferred from a local analysis.

The pronounced mutational robustness of native let-7 sequence-structure pairs against multiple-point mutations
might play a role in genetic robustness on a population level. The presence of native sequences in a population
allows for significant sequence variation within a species while still preserving the phenotype. This would
suggest that evolution is not identifying sequences that are locally robust but robust within entire
regions of sequence space.

An evolutionary analysis of the $\text{IFR}_{\sigma,k}^S$-spectrum of let-7 miRNAs shows that the $\text{IFR}_{\sigma,k}^S$-spectrum resembles phylogenetic
relationships. Statistical tests exhibit, that closely related species tend to have similar $\text{IFR}_{\sigma,k}^S$-spectra while
distant species tend to exhibit distinct $\text{IFR}_{\sigma,k}^S$-spectra. In general, higher level animal species have
larger $\text{IFR}_{\sigma,k}^S$-values than lower level animal species. The increased mutational robustness in higher level animals
appears to reflect the complexity of the organisms. 

The role of structure in the context of the $\text{IFR}_{\sigma,k}^S$-spectrum, differs substantially from the role of neutrality
in \cite{borenstein2006direct}: structure is not merely used to measure the effect of the mutation, but also affects how sequences
mutate. This constitutes effectively a feed-back loop between sequence and structure and is arguably the driving
factor enhancing the signal in native sequence-structure pairs.

We use a discrete metric for measuring the structural change induced by mutation in the definition of
$\text{IFR}_{\sigma,k}^S$-spectrum, instead of the base pair distance. This metric produces stable data: the standard
error of $\text{IFR}_{\sigma,k}^S$ by sampling $5000$ times is small, $\pm0.0065$, compared to the difference
of $\text{IFR}_{\sigma,k}^S$-spectra between native and control sequence-structure pairs and the difference of
$\text{IFR}_{\sigma,k}^S$-spectra across species, respectively. This allows us to analyze efficiently mutational robustness
against multiple-point mutations by Boltzmann sampling, obsoleting the need for large sample sizes.

In our analysis, we use the MFE structures, predicted by RNA minimum free-energy folding algorithm, as phenotypes.
The study can be enhanced by passing from MFE structures to partition functions of structures with respect to a
fixed sequence \cite{mccaskill1990equilibrium}. One can envision an analysis that considers the both: the dual partition
function, considered here and in addition the partition function of structures.

Traditionally, the ``information'' of a sequence is being identified with its actual sequence of nucleotides.
This perspective results in employing sequence alignments to quantify sequence similarities, the underline assumption
being, that similar sequences should have close biological relevant. This however is not entirely correct. 
Fontana {\it et al.} \cite{huynen1996smoothness} study the ruggedness of genotype to phenotype maps and show that similar
sequences can exhibit distinctly different phenotype. Furthermore, in the context of sequence design and detection of new
functional genes, sequences that fold into the same structure are considered to be equal \cite{lorenz2011viennarna,busch2006info}.
The presence of neutral networks
\cite{Reidys:97b,Gruner:96a,Gruner:96b} of RNA secondary structures, shows, that there are complimentary sequences
that fold into the same structure.
In other words, neither does sequence similarity necessarily imply phenotypic similarity nor does sequence dissimilarity
imply phenotypic dissimilarity. This motivates to augment the sequence information by incorporating additional factors. 

The analysis of $\text{IFR}_{\sigma,k}^S$-spectra represents such an augmentation: the let-7 sequences-structure pairs exhibit a distinctive
difference between native and random pairs. As a result, integrating the information of sequences and structures facilitates
at minimum the extension of the local analysis conducted in \cite{borenstein2006direct} and may as well, as a novel paradigm alone, lead to further
biological insights.

\section{FUNDING}
This research is partially funded by Thermo Fisher and the last author is a Thermo Fisher Scientific Fellow
in Advanced Systems for Information Biology.

\section{ACKNOWLEDGEMENTS}

Special thanks to Stanley Hefta for his input on this manuscript. 
We gratefully acknowledge the help of Kevin Shinpaugh and the computational support team at BI, 
Mia Shu, Thomas Li, Henning Mortveit, Madhav Marathe and Reza Rezazadegan for discussions.
The fourth author is a Thermo Fisher Scientific Fellow in Advanced Systems for Information Biology and
acknowledges their support of this work.

\subsubsection{Conflict of interest statement.} None declared.

%\bibliographystyle{NAR-natbib}
%\bibliography{NAR}

\begin{thebibliography}{10}

\bibitem{gu2003role}
Gu, Z., Steinmetz, L.~M., Gu, X., Scharfe, C., et al. (2003)
Role of duplicate genes in genetic robustness against null mutations.
{\em Nature,} {\bf 421}(6918), 63.

\bibitem{de2003perspective}
de~Visser, J. A. G.~M., Hermisson, J., Wagner, G.~P., Meyers, L.~A.,
  Bagheri-Chaichian, H., Blanchard, J.~L., Chao, L., Cheverud, J.~M., Elena,
  S.~F., Fontana, W., et al. (2003)
Perspective: evolution and detection of genetic robustness.
{\em Evolution,} {\bf 57}(9), 1959--1972.

\bibitem{schlichting1998phenotypic}
Schlichting, C.~D., Pigliucci, M., et al. (1998)
Phenotypic evolution: a reaction norm perspective.,
Sinauer Associates Incorporated, .

\bibitem{borenstein2006direct}
Borenstein, E. and Ruppin, E. (2006)
Direct evolution of genetic robustness in microRNA.
{\em Proceedings of the National Academy of Sciences,} {\bf 103}(17),
  6593--6598.

\bibitem{rodrigo2012describing}
Rodrigo, G. and Fares, M.~A. (2012)
Describing the structural robustness landscape of bacterial small RNAs.
{\em BMC evolutionary biology,} {\bf 12}(1), 52.

\bibitem{darnell2011rna}
Darnell, J.~E. (2011)
RNA: life's indispensable molecule,
Cold Spring Harbor Laboratory Press, .

\bibitem{breaker1996engineered}
Breaker, R.~R. (1996)
Are engineered proteins getting competition from RNA?.
{\em Current Opinion in Biotechnology,} {\bf 7}(4), 442--448.

\bibitem{serganov2007ribozymes}
Serganov, A. and Patel, D.~J. (2007)
Ribozymes, riboswitches and beyond: regulation of gene expression without
  proteins.
{\em Nature reviews. Genetics,} {\bf 8}(10), 776.

\bibitem{breaker1994inventing}
Breaker, R.~R. and Joyce, G.~F. (1994)
Inventing and improving ribozyme function: rational design versus iterative
  selection methods.
{\em Trends in biotechnology,} {\bf 12}(7), 268--275.

\bibitem{Waterman:78a}
Waterman, M.~S. (1978)
Secondary structure of single-stranded nucleic acids.
{\em Adv.\ Math.\ (Suppl.\ Studies),} {\bf 1}, 167--212.

\bibitem{Mathews:99}
Mathews, D., Sabina, J., Zuker, M., and Turner, D. (1999)
Expanded sequence dependence of thermodynamic parameters improves prediction of
  \textsc{RNA} secondary structure.
{\em J. Mol. Biol.,} {\bf 288}, 911--940.

\bibitem{Zuker:81}
Zuker, M. and Stiegler, P. (1981)
Optimal computer folding of larger {RNA} sequences using thermodynamics and
  auxiliary information.
{\em Nucleic Acids Res.,} {\bf 9}, 133--148.

\bibitem{Hofacker:94a}
Hofacker, I.~L., Fontana, W., Stadler, P.~F., Bonhoeffer, L.~S., Tacker, M.,
  and Schuster, P. (1994)
Fast Folding and Comparison of {RNA} Secondary Structures.
{\em Monatsh.\ Chem.,} {\bf 125}, 167--188.

\bibitem{lorenz2011viennarna}
Lorenz, R., Bernhart, S.~H., Zu~Siederdissen, C.~H., Tafer, H., Flamm, C.,
  Stadler, P.~F., and Hofacker, I.~L. (2011)
ViennaRNA Package 2.0.
{\em Algorithms for Molecular Biology,} {\bf 6}(1), 26.

\bibitem{gruber2008strategies}
Gruber, A.~R., Bernhart, S.~H., Hofacker, I.~L., and Washietl, S. (2008)
Strategies for measuring evolutionary conservation of RNA secondary structures.
{\em BMC bioinformatics,} {\bf 9}(1), 122.

\bibitem{kimura1983neutral}
Kimura, M. (1983)
The neutral theory of molecular evolution,
Cambridge University Press, .

\bibitem{mimouni2008analysis}
Mimouni, N.~K., Lyngs{\o}, R.~B., Griffiths-Jones, S., and Hein, J. (2008)
An analysis of structural influences on selection in RNA genes.
{\em Molecular biology and evolution,} {\bf 26}(1), 209--216.

\bibitem{price2011neutral}
Price, N., Cartwright, R.~A., Sabath, N., Graur, D., and Azevedo, R.~B. (2011)
Neutral evolution of robustness in Drosophila microRNA precursors.
{\em Molecular biology and evolution,} {\bf 28}(7), 2115--2123.

\bibitem{pollard2006rna}
Pollard, K.~S., Salama, S.~R., Lambert, N., Lambot, M.-A., Coppens, S.,
  Pedersen, J.~S., Katzman, S., King, B., Onodera, C., Siepel, A., et al.
  (2006)
An RNA gene expressed during cortical development evolved rapidly in humans.
{\em Nature,} {\bf 443}(7108), 167--172.

\bibitem{beniaminov2008distinctive}
Beniaminov, A., Westhof, E., and Krol, A. (2008)
Distinctive structures between chimpanzee and humanin a brain noncoding RNA.
{\em RNA,} {\bf 14}(7), 1270--1275.

\bibitem{pollard2006forces}
Pollard, K.~S., Salama, S.~R., King, B., Kern, A.~D., Dreszer, T., Katzman, S.,
  Siepel, A., Pedersen, J.~S., Bejerano, G., Baertsch, R., et al. (2006)
Forces shaping the fastest evolving regions in the human genome.
{\em PLoS genetics,} {\bf 2}(10), e168.

\bibitem{barrett2017sequence}
Barrett, C., Huang, F.~W., and Reidys, C.~M. (2017)
Sequence--structure relations of biopolymers.
{\em Bioinformatics,} {\bf 33}(3), 382--389.

\bibitem{lee1993c}
Lee, R.~C., Feinbaum, R.~L., and Ambros, V. (1993)
The C. elegans heterochronic gene lin-4 encodes small RNAs with antisense
  complementarity to lin-14.
{\em Cell,} {\bf 75}(5), 843--854.

\bibitem{tanzer2004molecular}
Tanzer, A. and Stadler, P.~F. (2004)
Molecular evolution of a microRNA cluster.
{\em Journal of molecular biology,} {\bf 339}(2), 327--335.

\bibitem{Nussinov:78}
Nussinov, R., Piecznik, G., Griggs, J.~R., and Kleitman, D.~J. (1978)
Algorithms for Loop Matching.
{\em SIAM J. Appl. Math.,} {\bf 35}(1), 68--82.

\bibitem{Turner:10}
Turner, D. and Mathews, D.~H. (2010)
{NNDB}: the nearest neighbor parameter database for predicting stability of
  nucleic acid secondary structure.
{\em Nucl. Acids Res.,} {\bf 38(Database)}, 280--282.

\bibitem{mccaskill1990equilibrium}
McCaskill, J.~S. (1990)
The equilibrium partition function and base pair binding probabilities for RNA
  secondary structure.
{\em Biopolymers,} {\bf 29}(6-7), 1105--1119.

\bibitem{huang2017efficient}
Huang, F.~W., He, Q., Barrett, C., and Reidys, C.~M. (2017)
An efficient dual sampling algorithm with Hamming distance filtration.
{\em arXiv preprint arXiv:1711.10549,}.

\bibitem{kozomara2013mirbase}
Kozomara, A. and Griffiths-Jones, S. (2013)
miRBase: annotating high confidence microRNAs using deep sequencing data.
{\em Nucleic acids research,} {\bf 42}(D1), D68--D73.

\bibitem{garcia2016rnadualpf}
Garcia-Martin, J.~A., Bayegan, A.~H., Dotu, I., and Clote, P. (2016)
RNAdualPF: software to compute the dual partition function with sample
  applications in molecular evolution theory.
{\em BMC bioinformatics,} {\bf 17}(1), 424.

\bibitem{levin2012global}
Levin, A., Lis, M., Ponty, Y., O'Donnell, C.~W., Devadas, S., Berger, B., and
  Waldisp{\"u}hl, J. (2012)
A global sampling approach to designing and reengineering RNA secondary
  structures.
{\em Nucleic acids research,} {\bf 40}(20), 10041--10052.

\bibitem{bonnet2004evidence}
Bonnet, E., Wuyts, J., Rouz{\'e}, P., and Van~de Peer, Y. (2004)
Evidence that microRNA precursors, unlike other non-coding RNAs, have lower
  folding free energies than random sequences.
{\em Bioinformatics,} {\bf 20}(17), 2911--2917.

\bibitem{freilich2010decoupling}
Freilich, S., Kreimer, A., Borenstein, E., Gophna, U., Sharan, R., and Ruppin,
  E. (2010)
Decoupling environment-dependent and independent genetic robustness across
  bacterial species.
{\em PLoS computational biology,} {\bf 6}(2), e1000690.

\bibitem{letunic2006interactive}
Letunic, I. and Bork, P. (2006)
Interactive Tree Of Life (iTOL): an online tool for phylogenetic tree display
  and annotation.
{\em Bioinformatics,} {\bf 23}(1), 127--128.

\bibitem{letunic2016interactive}
Letunic, I. and Bork, P. (2016)
Interactive tree of life (iTOL) v3: an online tool for the display and
  annotation of phylogenetic and other trees.
{\em Nucleic acids research,} {\bf 44}(W1), W242--W245.

\bibitem{federhen2011ncbi}
Federhen, S. (2011)
The NCBI taxonomy database.
{\em Nucleic acids research,} {\bf 40}(D1), D136--D143.

\bibitem{huynen1996smoothness}
Huynen, M.~A., Stadler, P.~F., and Fontana, W. (1996)
Smoothness within ruggedness: the role of neutrality in adaptation.
{\em Proceedings of the National Academy of Sciences,} {\bf 93}(1), 397--401.

\bibitem{busch2006info}
Busch, A. and Backofen, R. (2006)
INFO-RNA?a fast approach to inverse RNA folding.
{\em Bioinformatics,} {\bf 22}(15), 1823--1831.

\bibitem{Reidys:97b}
Reidys, C.~M., Stadler, P.~F., and Schuster, P. (1997)
Generic properties of combinatory maps and neutral networks of {RNA} secondary
  structures。.
{\em Bull. Math. Biol.,} {\bf 59}(2), 339--397.

\bibitem{Gruner:96a}
Gr\"{u}ner, R., Giegerich, R., Strothmann, D., Reidys, C.~M., Weber, J.,
  Hofacker, I.~L., Stadler, P.~F., and Schuster, P. (1996)
Analysis of {RNA} sequence structure maps by exhaustive enumeration {I.}
  structures of neutral networks and shape space covering..
{\em Chem. Mon.,} {\bf 127}, 355--374.

\bibitem{Gruner:96b}
Gr\"{u}ner, R., Giegerich, R., Strothmann, D., Reidys, C.~M., Weber, J.,
  Hofacker, I.~L., Stadler, P.~F., and Schuster, P. (1996)
Analysis of {RNA} sequence structure maps by exhaustive enumeration {II.}
  structures of neutral networks and shape space covering..
{\em Chem. Mon.,} {\bf 127}, 375--389.

\end{thebibliography}

\end{document}